\documentclass[sigconf]{acmart}
\usepackage{subcaption}
\usepackage{graphicx}
\usepackage{array}
\usepackage{makecell}
\usepackage{multirow}
\usepackage{xcolor}
\usepackage{capt-of}
\usepackage{booktabs}
\usepackage{tabularx}
\usepackage{enumitem}
\usepackage{adjustbox}
\usepackage{float}
\usepackage{stfloats}
\usepackage{cuted}
\usepackage{afterpage}
\AtBeginDocument{%
  }

\setcopyright{acmlicensed}
\copyrightyear{2026}
\acmYear{2026}
\acmDOI{XXXXXXX.XXXXXXX}
\acmConference[Conference acronym 'XX]{Make sure to enter the correct
  conference title from your rights confirmation email}{June 03--05,
  2018}{Woodstock, NY}
\acmISBN{978-1-4503-XXXX-X/2018/06}

\begin{document}

\title{Enhancing Relation Modeling with Social Attributes for Social Media Popularity Prediction}


\author{Bolun Zheng}
\author{Yuhao Luo}
\author{Wei Zhu}
\affiliation{%
  \institution{Hangzhou Dianzi University}
  \city{Hangzhou}
  \country{China}
}
\email{blzheng@hdu.edu.cn}
\email{242060267@hdu.edu.cn}
\email{242050252@hdu.edu.cn}

\author{Ning Xu}
\author{Anan Liu}
\affiliation{%
  \institution{Tianjin University}
  \city{Tianjin}
  \country{China}
}
\email{ningxu@tju.edu.cn}
\email{anan0422@gmail.com}

\author{Lingyu Zhu}
\affiliation{%
  \institution{City University of Hong Kong}
  \city{Hong Kong}
  \country{China}
}
\email{lingyzhu-c@my.cityu.edu.hk}

\author{Canjin Wang}
\affiliation{%
  \institution{Xinhua Zhiyun Technology Co., Ltd.}
  \city{Hong Kong}
  \country{China}
}
\email{Canjin Wang@shuwen.com}

\renewcommand{\shortauthors}{Trovato et al.}

\begin{abstract}
Recent studies highlight the critical role of retrieval-augmented mechanisms in social media popularity prediction (SMPP). 
Although such frameworks have improved SMPP performance by leveraging historical posts, existing methods still suffer from the low retrieval accuracy due to the oversight of relative relationships among UGC instances.
To address this limitation, we propose a novel Relation-Enhanced Retrieval-Augmented framework (RE-Rag) that models UGC similarity as a continuous relation jointly driven by semantic content and social attributes.  
Specifically, RE-Rag employs a Semantic-Attribute Retriever (SAR) to obtain instances aligned in both semantic and social-attribute distributions.
Subsequently, we design a Relation-Guided Predictor (RGP): first, cross-attention encodes multimodal features of retrieved instances; then, a relative relation graph is introduced to guide attention weight allocation, forming a Relation-Guided Transformer (RGTs) that dynamically modulate attention weights based on relative attribute relations to capture the interplay between semantics and various social attributes. The refined features are fused with the target instance for popularity prediction.
Experiments on three public benchmarks show that RE-Rag consistently outperforms state-of-the-art methods in both prediction accuracy and retrieval efficiency. The code and data are available at \url{https://github.com/BBEC-opt/RE_RAG}.
\end{abstract}

\ccsdesc[500]{Information systems → Multimedia information systems.}

\keywords{Multimedia popularity, retrieval augmentation}

\maketitle

\section{Introduction}
\begin{figure}[t]
    \centering
    \includegraphics[width=0.47\textwidth]{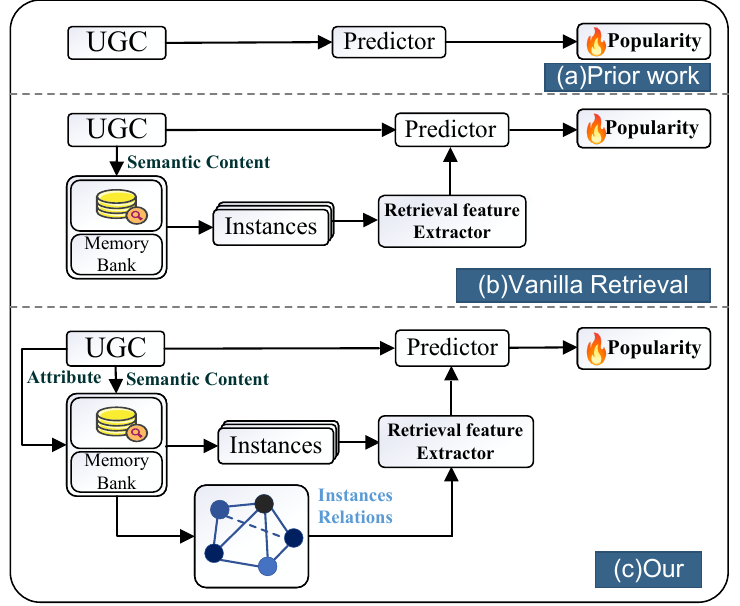}
    \caption{Comparison among (a) non-retrieval models, (b) vanilla retrieval models, and (c) our proposed model.}
    \label{fig:1}
\end{figure}
Social media platforms have enjoyed effective growth over the past decade. 
Every day, hundreds of millions of users create information on various social media platforms. 
Predicting potential popularity of user-generated content (UGC) before its distribution is essential for platforms to optimize resource allocation \cite{Storage1,Storage2}, improve advertising strategies \cite{Advertising1,Advertising2}, refine marketing approaches \cite{market1,market2}, develop effective policies \cite{police, suhaimin2023social}, and content ecosystem governance \cite{social,social2}.
Due to the high potential commercial value, the social media popularity prediction (SMPP) has emerged as a critical research topic and attracts enormous academic attention \cite{newman2025digital}. 
Early approaches (Figure~1(a)) extract diverse features from content, metadata, and social interactions, then use statistical models or neural networks to directly predict popularity scores \cite{khosla2014makes,chen2016micro,cho2024amps,pinto2013using}. 
However, such models typically treated each post in isolation and relied on limited contextual information, which hindered their ability to capture underlying user behavior trends and patterns of information diffusion. 
To address this limitation, recent studies (Figure~1(b)) introduce retrieval-enhanced frameworks by leveraging similar historical posts as auxiliary features for popularity prediction \cite{MMRA,NIPA}.

Despite their effectiveness, existing retrieval-enhanced methods often equate semantic similarity with instance relevance, overlooking the fact that semantically similar UGCs may exhibit different popularity patterns. Such discrepancies arise because similar content can be associated with different social contexts and propagation processes, leading to diverse popularity outcomes.
Even highly similar posts may be embedded in very different social contexts, exposed to different user groups, and activated through distinct diffusion pathways, leading to drastically different popularity outcomes\cite{cascades}.


Beyond semantic-based retrieval, some studies \cite{RAGtrans,skapp} utilize social attributes to enhance the RAG. However, they mainly treat social attributes as auxiliary cues for similarity of contents, without explicitly considering their role as indicators of the underlying distribution mechanism. Thus, they struggle to capture the subtle yet crucial distinctions among UGCs.


In this work, we rethink the mechanism of social media communication and propose a dual-factor model, where social media popularity is
governed by two independent factors--content quality and distribution mechanism. The former reflects how engaging the content itself is, while the latter characterizes how many users the platform exposes the content
to, determined by social attributes (author, category, language,
etc.).

Based on this perspective, we propose a relation-enhanced retrieval augmentation framework (RE-Rag) that model relative attribute relations among UGC instances for effective instance retrieval and accurate popularity prediction.
Specifically, we first introduce a Semantic-Attribute Retriever (SAR) to retrieve valuable UGC instances from the database basing on semantic cues and attribute-based matching score, and build a relation map of the target UGC and corresponding retrieved instances.
Then two relation-guided transformers (RGTs) are involved, constructing a relation-guided predictor (RGP) to predict popularity score using the target UGC and retrieved instances with the assist of the relation map obtained in SAR.
Generally, our major contributions can be summarized as follows:
\setlength{\leftmargini}{0pt}
\begin{itemize}
\item We propose a relation-enhanced retrieval augmentation framework namely RE-Rag, that models UGC similarity as a continuous relation jointly driven by semantic content and social attributes to enhance popularity prediction.  
\item We design two core modules, semantic-attribute retriever and relation-guided predictor, against distinguishing valuable UGCs based on relation modeling driven by social attributes for retrieval augmentation.
\item Extensive experiments on three public benchmarks demonstrate that the proposed relation modeling strategy is effective in retrieval stage, and our RE-Rag achieves state-of-the-art performance with better retrieval efficiency.

\end{itemize}

\section{RELATED WORK}
\subsection{Popularity Prediction}
Social Media Popularity Prediction aims to predict the popularity of UGC based on multimodal information and has shown practical value across domains. Early approaches relied on predefined features and constructed predictive rules using statistical methods such as linear regression \cite{szabo2010predicting}, but these methods struggled to capture complex nonlinear relationships. Subsequently, researchers combined handcrafted features (e.g., content attributes, user profiles) with machine learning models \cite{SVR,MFTC,MFTM,lai2020hyfea} to characterize diffusion patterns. However, such methods remain sensitive to feature quality, required extensive domain expertise, and lack scalability.

In recent years, deep neural networks have driven the development of end-to-end representation learning, enabling the automatic discovery of complex patterns from data and offering stronger feature modeling and scalability in SMPP tasks. For example, VSCNN \cite{VSCNN} employs convolutional neural networks to jointly model visual content and social context features, DTCN \cite{DTCN} introduces temporal coherence by learning continuous temporal context for predicting popularity dynamics, CBAN \cite{CABN} enhances multimodal feature interaction and expressiveness through bipolar attention, MASSL \cite{MASSL} proposes a multimodal prediction framework that combines feature extraction strategies with variational autoencoders, HMMVED \cite{HMMVED} adopts a hierarchical multimodal variational encoding–decoding framework to capture cross-modal semantic structures, Liao et al. \cite{liao2019popularity} integrate article content with temporal features via deep fusion and hierarchical feature extraction strategies to improve prediction performance, and TGANN \cite{qian2022popularity} employs text-guided attention for multimodal fusion.

Nevertheless, most existing approaches focus on single-instance popularity prediction, relying on limited modal cues. This hinders capturing broader semantic associations and diffusion patterns, thus limiting performance in dynamic social media environments.

\subsection{Retrieval-Augmented Generation}
Retrieval-augmented generation (RAG) is a paradigm that integrates neural network models with external retrieval modules, enabling the model to dynamically acquire and incorporate relevant information from large-scale knowledge bases, thereby improving both prediction and generation performance \cite{gao2023retrieval,lewis2020retrieval}. This paradigm has been widely applied in domains such as large language models \cite{Lewis2020RAG,Borgeaud2022RETRO}, recommender systems \cite{Zhang2021RARec}, and social network analysis \cite{scrag}. Its effectiveness has also been validated in the task of SMPP. For example, AFRF \cite{AFRF} retrieves similar popularity time series through angular features, while MMRA and NIPA \cite{MMRA,NIPA} leverage multimodal semantic content retrieval to obtain similar instances and extract multimodal features for popularity prediction. However, these approaches often overlook social relation information, which may lead to spurious correlations between retrieved and target instances, thereby limiting the quality of the extracted knowledge.


Recently, RAGTrans \cite{RAGtrans} clusters UGC contents into virtual attributes and aggregates them with social attributes through a hypergraph \cite{ragraph}, while SKAPP \cite{skapp} enhances retrieval with BM25 algorithm and  Selective Refiner. Although both leverage social attributes, they regard them as auxiliary cues for sample retrieval. RAGTrans further incorporates social attributes into hypergraph modeling, but the predefined aggregation only captures coarse associations and fails to characterize the underlying distribution mechanism. Therefore, we argue that social attributes should serve as proxy variables for modeling platform distribution mechanisms rather than auxiliary features for content similarity measurement.


\section{METHODOLOGY}
\begin{figure*}[t] 
    \centering
    \includegraphics[width=\textwidth]{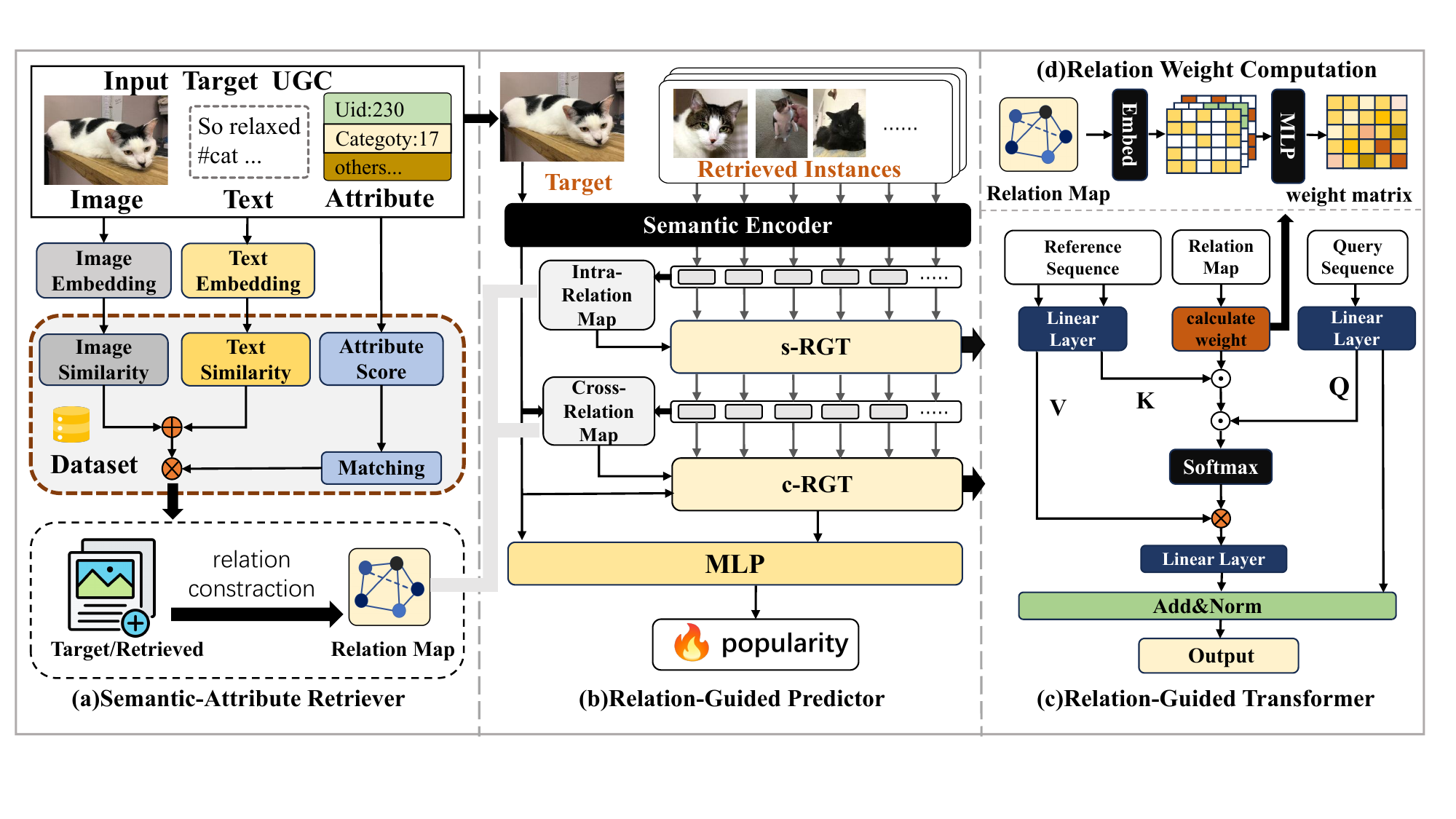} 
    \caption{Overview of the proposed RE-Rag framework:
(a) Semantic-Attribute Retriever combines multimodal semantic similarity with attribute compensation to select relevant samples.
(b) Relation-Guided Predictor, comprising a Semantic Encoder and Relation-Guided Transformer, extracts retrieved sequence features for prediction.
(c) Relation-Guided Transformer embeds relative relation patterns into attention to highlight critical features.
(d) Relation Weight Computation Details.}
    \label{fig:main} 
\end{figure*}

In this section, we present the proposed RE-Rag framework (Figure~\ref{fig:main}). Given an input UGC instance, a Semantic-Attribute Retriever first identifies relevant references. These instances are then processed by a retrieval feature extractor, comprising a semantic encoder and a relation-guided transformer, to derive informative representations for popularity prediction. Detailed descriptions of each component are provided in the following subsections.
\subsection{Problem Definition}
The goal of SMPP is to predict the popularity of UGCs using their content and social attributes before they actually communicate on the platform.
Generally, we adopt normalized view counts as the popularity \cite{xu2025smtpd}, which can be written as:
\begin{equation}
p = \log_2\left(\frac{v}{d} + 1\right),
\end{equation}
where $p$ denotes UGC's popularity score, $v$ denotes its view count, and $d$ denotes the number of days since its communication.
This definition is designed to eliminate the influence of time span on view growth, thereby providing a fairer assessment of UGC popularity on a unified time scale.

\subsection{Semantic-Attribute Retriever}

In this section, we propose a Semantic-Attribute Retriever, aiming at retrieving instances with good consistency of popularity. 
The UGC's popularity is affected by both content semantics and social attributes.
Therefore, the proposed SAR involves both of them to construct the database and multi-modal retriever.

\subsubsection{\textbf{Database Construction}}

We constructed a multimodal retrieval database to retrieve UGC instances with similar popularity. 
Each UGC entry is represented as <$content$, $attribute$, $label$>.
Specifically, $content=[\boldsymbol{f}_{text},\boldsymbol{f}_{Img}] $ denotes vector representations derived from text features $f_{text}$ and image features $\boldsymbol{f}_{Img}$ through pre-trained models. 
This $\text{label}$ indicates the corresponding popularity.
This $attribute$ represents the social attributes of a UGC, including the category, author information, and language. Detailed definitions of social attributes across different datasets are provided in the supplementary material.

\subsubsection{\textbf{Multimodal Retrieval}}
To retrieve UGC instances similar to the target, we design a multi-modal retriever that fully considers both content semantic and social attributes.
The retriever consists of two stages, semantic matching and attribute compensation.
In the semantic matching stage, we measure the textual and image semantic similarities between the target and candidate using the cosine similarity function:
\begin{eqnarray}
    s^{target}_{candi}\Big|_{Img} &= \text{sim}(\boldsymbol{f}^{target}_{Img},\boldsymbol{f}^{candi}_{Img})\\
     s^{target}_{candi}\Big|_{text} &= \text{sim}(\boldsymbol{f}^{target}_{text},\boldsymbol{f}^{candi}_{text})
\end{eqnarray}
where $\text{sim}(\cdot,\cdot)$ denotes the cosine similarity function, $\boldsymbol{f}^{target}_{*}$ and $\boldsymbol{f}^{candi}_{*}$ denote the features of the target UGC and candidate UGC.
Then we can have the semantic matching score as:
\begin{equation}
    s^{target}_{candi} = \alpha\cdot s^{target}_{candi}\Big|_{Img} + (1-\alpha)\cdot s^{target}_{candi}\Big|_{text}
\end{equation}
where $\alpha \in (0, 1)$ is a hyper-parameter that balances the contribution of the text modal and image modal.

In the attribute compensation stage, we introduce social attributes to compensate for the popularity gap among UGCs with similar semantics.
Social attributes are frequently associated with specific audiences and delivery pathways \cite{de2010birds}, as rarer attributes tend to carry more information and focusing on them helps distinguish propagation patterns.

Considering that there are $m$ social attributes $\boldsymbol{A}=\{a_1, a_2, \dots, a_m\}$ involved in the dataset (The details of the social attributes in different datasets are illustrated in the supplementary material), and each social attribute gets $m_i$ possible values, labeled as $\{u^{i}_{1}, u^{i}_{2}, \dots, u^{i}_{n_i}\}$, $i\in\ \mathbb{N}^{[1,m]}$, the social attributes of a UGC instance can be denoted as $<u^{1}_{l_1}, u^{2}_{l_2}, \dots, u^{m}_{l_m}>$, $l_{i}\in \mathbb{N}^{[1,n_i]}$.
Then we can calculate its compensation weight based on its rarity.
Specifically, we evaluate the rarity of each social attribute from global and local perspectives.
From the global sight, we adopt the inverse document frequency (IDF) \cite{sparck1972statistical_tf-idf} as the global rarity, written as:
\begin{equation}
\text{GR}(a_i) = \log_2 \frac{N_{total}}{\text{freq}(u^{i}_{l_i})+1} + 1
\label{eq:gr}
\end{equation}
where $N_{total}$ denotes the totals of UGCs in the database, $\text{freq}(u^{i}_{l_i})$ returns the occurred count of $u^{i}_{l_i}$ in the database.
From the local sight, we measure the local rarity within one social attribute following Eq.~\ref{eq:gr}, which can be expressed as:
\begin{equation}
\text{LR}(a_i) = \log_2 (\frac{\text{freq}_{\text{max}}(a_i)}{\text{freq}(u^{i}_{l_i})+1} + 1)
\end{equation}

\begin{equation}
\text{freq}_{\text{max}}(a_i) = \text{Max}(\text{freq}(u^{i}_{1}),\text{freq}(u^{i}_{2}),\ldots,\text{freq}(u^{i}_{n_i}))  
\end{equation}
Then, we can have the rarity of the social attribute $a_i$ by calculating the geometric average of $\text{GR}(a_i)$ and $\text{LR}(a_i)$, written as:
\begin{equation}
\text{R}(a_i) = \sqrt{\text{GR}(a_i) \cdot \text{LR}(a_i)}
\end{equation}
Subsequently, we introduce a matching function to calculate the compensation weight for a candidate UGC, expressed as:
\begin{equation}
w\Big|^{target}_{candi} = \prod^{m}_{i=1} (1 + \text{Norm}(\text{R}(a_i)) \cdot \text{T}(a_{i}^{candi},a_{i}^{target}))
\end{equation}
where Norm$(\cdot)$ denotes a min-max normalization, its detailed formulation is provided in the supplementary material, $a_{i}^{candi}$ and $a_{i}^{target}$ denotes the social attributes of the candidate UGC and target UGC, and $\text{T}(\cdot, \cdot)$ denotes a matching function, written as:
\begin{equation}
\small
\text{T}(x,y) =
\begin{cases}
0, & x = y  \\
1, &x \neq y
\end{cases}
\end{equation}
Finally, we can have the compensated matching score as:
\begin{equation}
\hat{s}\Big|^{target}_{candi} = s\Big|^{target}_{candi} \cdot w\Big|^{target}_{candi}
\end{equation}
where $\hat{s}\Big|^{target}_{candi}$ denotes the compensated matching score, which will finally be used to retrieve top-$N_{retrieval}$ similar UGC instances for the following popularity prediction.

\subsection{Relation-Guided Predictor}
The RGP is supposed to predict the popularity of the input UGC with the assistance of retrieved instances in the SAR.
As shown in Figure~\ref{fig:main}(b), we first introduce a semantic encoder to translate the input UGC and retrieved instances to feature vectors, respectively.
Then, two RGTs, the self-relation-guided transformer (s-RGT) and cross-relation-guided transformer (c-RGT) are sequentially introduced to refine the features of retrieved instances.
Finally, the refined features and the input UGC features are jointly fed into an MLP to predict popularity.

\subsubsection{\textbf{Semantic Encoder}}
Taking a UGC as input, the semantic encoder first uses pre-trained feature extractors to  translate the image content and textual content to corresponding feature vectors, respectively.
Assuming the translated image feature is $f_{Img}$ and the translated textual feature is $f_{text}$, we adopt a cross-attention mechanism \cite{vaswani2017attention} to fuse them, which can be expressed as:
\begin{align}
\boldsymbol{h}_{Img}  &= \text{softmax}\Big(\frac{\boldsymbol{f}_{Img} \, \cdot \boldsymbol{f}_{text}^\top}{\sqrt{d_k}}\Big) \, \boldsymbol{f}_{text}, \\
\boldsymbol{h}_{text} &= \text{softmax}\Big(\frac{\boldsymbol{f}_{text} \, \cdot \boldsymbol{f}_{Img}^\top}{\sqrt{d_k}}\Big) \, \boldsymbol{f}_{Img}, \\
\boldsymbol{z}        &= \langle \boldsymbol{h}_{Img}, \boldsymbol{h}_{text} \rangle,
\end{align}
where $d_k$ denotes the feature dimension used for cross-modal attention, $\langle \cdot, \cdot \rangle$ denotes the concatenate operation, and $z$ denotes the fused feature produced by the semantic encoder.

\subsubsection{\textbf{Relation Map with Social Attributes}}
Existing research has demonstrated that incorporating a relation map between the target and retrieved instances can facilitate feature optimization \cite{peng2024graph}. 
To explicitly model relative attribute relations within the attention mechanism, we construct a relation map that not only characterizes the attribute pattern correspondences between arbitrary instance pairs but also differentiates the attribute information of the target instance, thereby highlighting the key relations most relevant to the prediction task.  

Given a set of UGC instances, containing $N_{retrieval}$ retrieved UGC instances $\{I_{1}, I_{2}, \ldots, I_{N_{retrieval}}\}$ and the corresponding target $I_{target}$, for any two elements $I_i$ and $I_j$ in the collection, $i,j\in\{1,2,\ldots,N_{retrieval},target\}$, we can calculate their relation degree under the social attribute $a_k\in \boldsymbol{A}$ as:
\begin{equation}
\boldsymbol{E}_{a_k}(I_{i},I_{j}) =
\begin{cases}
0, & a_k|_{I_i} \neq a_k|_{I_j}, \\
1, & a_k|_{I_i} = a_k|_{I_j} \neq a_k|_{I_{target}}, \\
2, & a_k|_{I_i} = a_k|_{I_j} = a_k|_{I_{target}}.
\end{cases}
\end{equation}
where $a_k|_{I_i}$ denotes the value of the social attribute $a_k$ in the UGC instance $I_i$.
Thus, we can obtain a relation map for $a_k$ by traversing all possible combinations of $(I_i, I_j)$, denoted as $\boldsymbol{E}_{a_k}$.
Then, By stacking the $m$ attribute-wise relation maps, we can obtain the overall relation map $\boldsymbol{E}$.
We calculate two relation maps, the self-relation map $\boldsymbol{E}^{self}$ and the cross-relation map $\boldsymbol{E}^{cross}$, for the subsequent s-RGT and c-RGT respectively.
Specifically, the $I_{target}$ is only included in the calculation of For $\boldsymbol{E}^{cross}$, but not the For $\boldsymbol{E}^{self}$.
Therefore, the shape of $\boldsymbol{E}^{self}$ is $N_{retrieval}\times N_{retrieval} \times m$, while the shape of For $\boldsymbol{E}^{cross}$ is  $1\times N_{retrieval} \times m$.

\subsubsection{\textbf{Relation-Guided Transformer}}

Though the relation map is graph-like data, simply adopting graph convolution to handle the relation map may suffer from its limited receptive fields, especially when we get a large number of retrieved instances.
Against this limitation, we propose a relation-guided transformer to refine the feature sequence of retrieved instances from a global sight.

Considering a group of $<\boldsymbol{query}, \boldsymbol{key}, \boldsymbol{value}>$ group labeled as $<\boldsymbol{Q},\boldsymbol{K}, \boldsymbol{V}>$, we first introduce a weight matrix $M$ calculated by the relation map $\boldsymbol{E}$ to refine the $K$, then use the refined $\boldsymbol{K}$ to complete the rest of the calculation as a standard transformer does. 
Specifically, we use an embedding layer to obtain embeddings of $\boldsymbol{E}$, then introduce an MLP to further calculate $M$, which can be expressed as:
\begin{equation}
    \boldsymbol{M} = \text{MLP}(\text{Embed}(\boldsymbol{E}))
\end{equation}
where $\text{Embed}(\cdot)$ denotes the embedding operation. 
So that the refined $K$ can be written as:
\begin{equation}
    \boldsymbol{\hat{K}} = \boldsymbol{K}\cdot  \boldsymbol{M}
\end{equation}

In practice, we design two RGTs, self-relation-guided transformer (s-RGT) and cross-relation-guided transformer (c-RGT), that the s-RGT is supposed to conduct a preliminary refinement using the internal relationship among retrieved instances, while the c-RGT is supposed to conduct a further refinement using the cross relationship between the target and retrieved instances.
Given $N$ retrieved instances $\{\boldsymbol{z}_1, \boldsymbol{z}_2, \ldots, \boldsymbol{z}_N\}$, we concatenate each semantic feature $\boldsymbol{z}_i$ with its popularity $p_i$ and similarity score $s_i$, then map the concatenated vector through an MLP to obtain a fused representation $\tilde{\boldsymbol{z}}_i$. and then adding positional encodings \cite{wang2020position} to formulate the input of s-RGT $\boldsymbol{z}_{in}$, which can be expressed as:

\begin{equation}
    \tilde{\boldsymbol{z}}_i = \text{MLP}\big(<\boldsymbol{z}_i, p_i, \hat{s}\Big|^{target}_{i}>\big)
\end{equation}
\begin{equation}
\boldsymbol{z}_{\text{in}} = \{ \tilde{\boldsymbol{z}}_i + \text{PosEnc}(i) \}_{i=1}^{N_{retrieval}}
\end{equation}
where $\text{PosEnc}(\cdot)$ denotes the positional encoding operation, $p_i$ and $\hat{s}\Big|^{target}_{i}$ denote the corresponding popularity and compensated matching score of $z_i$.
For the s-RGT, we use three linear layers to generate $<\boldsymbol{Q}, \boldsymbol{K}, \boldsymbol{V}>$ from $\boldsymbol{z}_{in}$, and obtain a self-refined feature ${\boldsymbol{z}_{self}}$.
For the c-RGT, we use one linear layer to generate $\boldsymbol{Q}$ from $\boldsymbol{z}_{target}$ and another two linear layers to generate $<\boldsymbol{K},\boldsymbol{V}>$ from ${\boldsymbol{z}_{self}}$, where $\boldsymbol{z}_{target}$ denotes the features of the input UGC derived by the semantic encoder.
Thus, we can obtain the further refined feature ${\boldsymbol{z}_{cross}}$ through the c-RGT.

\subsubsection{Predictor}
In the final prediction stage, we combine the $\boldsymbol{z}_{target}$ and ${\boldsymbol{z}_{cross}}$ together, and send them to an MLP to predict the popularity of the input UGC, which can be expressed as:
\begin{equation}
    \hat{p} = \text{MLP}(<\boldsymbol{z}_{target}, \boldsymbol{z}_{cross}>)
\end{equation}
We measure the L2 loss between predicted popularity and ground-truth popularity, providing full supervision for our RE-Rag.

\section{EXPERIMENTS}
In this section, we evaluate the effectiveness of the proposed RE-Rag model. We conduct experiments on three datasets and compare it with multiple strong baselines. Ablation studies are performed to verify the contribution of key components. In addition, a series of experiments are carried out to further analyze the retrieval capability, prediction performance, and stability of the model.

\subsection{Implementation  Details}
\subsubsection{Datasets and Metrics}
We adopt three publicly available datasets for SMPP to evaluate our method, including two widely used ICIP \cite{icip} and SMPD (Image-300K) \cite{wu2024smp}, and a newly proposed SMTPD \cite{xu2025smtpd}.
The details of three dataset are illustrated in Table~\ref{tab:dataset_comparison}.
The dataset split ratio is 8:1:1 for training, validation, and test sets, respectively.
Since ICIP and SMTPD provide temporal popularity labels, we only predict the popularity of 30th-day as previous studies \cite{wu2024smp, tatar2014survey} did.
Across all datasets, we use category information at different hierarchical levels and author information as attributes, and for the SMTPD dataset, due to its multilingual content, we additionally incorporate language as an attribute.

\begin{table}[htbp]
\centering
\small  
\setlength{\tabcolsep}{1pt}  
\caption{Dataset statistics used in our experiments.}
\label{tab:dataset_comparison}
\begin{tabular*}{\linewidth}{@{\extracolsep{\fill}}lccccc}
\toprule
Name & \#Samples & \#Users & \#Lang. & Std (Pop.) & Source \\
\midrule
ICIP   & 20,337   & 17,302   & 1(EN)     & 1.54           & Flickr   \\
SMPD   & 305,595  & 38,307   & 1(EN)     & 2.47           & Flickr   \\
SMTPD  & 282,000  & 152,700  & 90+   & 4.15   & YouTube  \\
\bottomrule
\end{tabular*}
\end{table}

\subsubsection{\textbf{Settings and Metrics}}
We employ the Adam \cite{kingma2017adammethodstochasticoptimization} optimizer with an initial learning rate of 1e-5 and a batch size of 64 and a weight decay of 1e-4. The learning rate is further adjusted using PyTorch’s ReduceLROnPlateau scheduler at the end of each epoch.
An early stopping strategy is applied, terminating training if validation performance does not improve for 5 consecutive epochs. 
Each model is run five times, and all reported results are averaged over these runs for robustness.
We adopt three commonly used metrics: mean squared error (MSE), mean absolute error (MAE), and Spearman’s rank correlation coefficient (SRC) to fairly evaluate the performance of models compared.
Details of the feature embedding model are provided in the supplementary material.

\subsection{Main Results}
We evaluate the effectiveness of our model by comparing it with the following state-of-the-art approaches, including the feature engineering-based methods: Hyfea \cite{lai2020hyfea}; the deep learning-based methods: HMMVED \cite{HMMVED}, MASSL \cite{MASSL}, UHAN \cite{UHAN}, CBAN \cite{CABN}  Jab \cite{JAB}; the retrieval-augmented methods: MMRA \cite{MMRA}, ICPF \cite{ICPF}, RAGtrans \cite{RAGtrans}, SKAPP \cite{skapp}.
The comparison results are presented in Table~\ref{tab:performance}.
The proposed model outperforms the comparison methods on the vast majority of metrics.
Notably, our RE-Rag achieves relative improvements of $29.61\%, 29.43\%, 3.80\%$ in MSE, MAE and SRC over the second-best method on the SMPD dataset. 
Besides, our method also exhibits encouraging performance on the more challenging SMTPD dataset.
Although some comparison methods also employ retrieval-enhanced strategies, their performance gains remain limited due to constraints in retrieval quality and modeling approaches. 
SKAPP achieves competitive performance on the smaller ICIP dataset, where weaker distribution-related attributes (e.g., ispro and HasStats) and limited data scale reduce the difficulty of modeling propagation mechanisms, resulting in similar performance across methods. However, it exhibits clear degradation on the more complex SMPD and SMTPD datasets, which provide richer semantics, larger volumes, and attributes more closely correlated with distribution mechanisms. This indicates that its multi-stage retrieval and feature modeling strategies are insufficient to capture distribution-related propagation differences.
In contrast, RE-Rag consistently maintains strong performance across datasets of varying scales and semantic complexity, demonstrating superior robustness and generalization capability (see Tables~\ref{tab:comparison} and \ref{tab:re}).

\begin{table*}[!t]
\caption{Performance comparison on three real-world datasets. The best results are in \textbf{bold} and the second-best are \underline{underlined}. Lower values of MSE and MAE, and higher values of SRC, indicate better performance.}
\label{tab:performance}
\centering
\normalsize
\setlength{\tabcolsep}{4pt} 
\renewcommand{\arraystretch}{1.1}
\begin{tabular*}{\textwidth}{@{\extracolsep{\fill}} l l *{9}{c} }
\toprule
 & \multicolumn{1}{c}{}&
 \multicolumn{3}{c}{\textbf{ICIP}} & \multicolumn{3}{c}{\textbf{SMPD}} & \multicolumn{3}{c}{\textbf{SMTPD}} \\
\cmidrule(lr){3-5} \cmidrule(lr){6-8} \cmidrule(lr){9-11}
& \textbf{Pub}
 & \textbf{MSE} & \textbf{MAE} & \textbf{SRC} & \textbf{MSE} & \textbf{MAE} & \textbf{SRC} & \textbf{MSE} & \textbf{MAE} & \textbf{SRC} \\
\midrule
HyFea &MM'20 & 2.0788 & 1.0460 & 0.3842 & 1.9032 & 0.9316 & 0.8324 & 8.0977 & 2.1585 & 0.6808 \\
HMMVED &IEEE TMM’23 & 2.0588 & 1.0328 & 0.4379 & 2.3585 & 1.1100 & 0.7866 & 6.2550 & 1.9414 & 0.7768 \\
MASSL &ICCIP'22 & 2.0621 & 0.9565 & 0.4602 & 1.7514 & 0.9045 & 0.8435 & 7.6974 & 2.1031 & 0.7062 \\
UHAN  &WWW'18 & 2.1045 & 1.1047 & 0.5731 & 4.7491 & 1.6275 & 0.6698 & 7.2069 & 2.0726 & 0.6793 \\
CBAN  &NEUCOM'22 & 1.7898 & 0.9239 & 0.4935 & 1.6421 & 0.8661 & 0.8560 & 7.2139 & 2.0534 & 0.7206 \\
Jab   &ICWSM'22 & 2.2529 & 1.0845 & 0.4047 & 2.3673 & 1.0293 & 0.7862 & 9.7532 & 2.4804 & 0.6009 \\
MMRA  &SIGIR'24 & 1.7549 & 0.8978 & 0.5049 & 1.6023 & 0.8455 & 0.8621 & 6.2997 & 1.8974 & 0.7803 \\
ICPF  &AAAI'25 & 1.9031 & 0.9021 & 0.3874 & 2.7408 & 1.1589 & 0.7329 & 7.4978 & 1.9850 & 0.7327 \\
RAGtrans &KDD'24 & 1.5719 & 0.8332 & 0.6387 & \underline{1.0318} & 0.6541 & 0.8927 & \underline{5.2274} & \underline{1.6516} & \underline{0.8125} \\
SKAPP &AAAI'25 & \textbf{1.1075} & \underline{0.6829} & \textbf{0.6975} & 1.1926 & \underline{0.6487} & \underline{0.9012} & 7.4017 & 2.0131 & 0.7796 \\
\midrule[0.5pt]
RE-Rag (Ours) & & \underline{1.1360} & \textbf{0.6617} & \underline{0.6740} & \textbf{0.7263} & \textbf{0.4577} & \textbf{0.9355} & \textbf{4.2287} & \textbf{1.4697} & \textbf{0.8338} \\

\bottomrule
\end{tabular*}
\end{table*}
\subsection{Ablation Study}
We conduct extensive ablation studies to assess the contribution of each key component in RE-Rag. The effects of individual social attributes are reported in the supplementary material.
\subsubsection{\textbf{Impact of Semantic-Attribute Retriever.}}
We evaluated two variant models:
(1) \textbf{w/o Retrieval}: the retrieval mechanism was completely removed; 
(2) \textbf{w/o Attribute}: During the retrieval phase, attribute compensation scores are eliminated, and the process relies solely on semantic similarity; 
(3) \textbf{w/o LR}: Only Use GR; 
(4) \textbf{w/o GR}: Only Use LR; 
The results are shown in Table~\ref{tab:ablation}.

From the results, removing the retrieval mechanism leads to a sharp performance reduction, which confirms the effectiveness of retrieval augmentation.
Besides, eliminating attribute compensation scores also weakens the overall performance, high similarity in semantic space alone does not guarantee consistency in propagation mechanisms. Without attribute-based compensation, the number of pseudo-correlated instances increases, thereby impairing overall performance.
Additionally, we find that removing either local rarity or global rarity during retrieval leads to a performance drop, indicating that the two play complementary roles in candidate selection. Local rarity highlights the distinctiveness of an instance within its attribute space, while global rarity reflects the overall information value of the attribute across the dataset. Their combination not only mitigates redundancy caused by high-frequency attribute values but also preserves low-frequency yet representative features, thereby enhancing the effectiveness of retrieval augmentation.

\subsubsection{\textbf{Impact of Semantic Encoder.}} To evaluate the contribution of each feature component in the semantic encoder to overall performance, we designed three investigation models: (1) \textbf{w/o Text}: textual modality removed; (2) \textbf{w/o Image}: image modality removed; (3) \textbf{w/o Cross}: cross-attention mechanism removed. 

Removing either the textual feature or the image feature would bring a visible performance reduction. 
The absence of the textual feature is more serious than the image feature.
This suggests that the textual feature could provide more accurate semantic information compared to the image feature. 
In addition, the cross-attention mechanism also contributes to obtaining more effective feature representation by fusing the textual and image features.

\subsubsection{\textbf{Impact of the Relation-Guided Transformer.}} In this part, we evaluate the contribution of components related to the RGTs.
Specifically, we design three experimental models:
(1) \textbf{w/o Relation Map}: removing the relation map from the RGTs, which means replacing the RGTs with a standard transformer; 
(2) \textbf{w/o s-RGT}: removing the s-RGT; 
(3) \textbf{w/o c-RGT}: removing the c-RGT.

As shown in Table~\ref{tab:ablation}, removing the relation map leads to a noticeable performance drop, indicating that incorporating a relation-guided mechanism during context feature extraction helps the model more effectively identify valuable information from retrieved instances. Similarly, removing either s-RGT or c-RGT also degrades performance, particularly increasing MAE, which suggests that constructing a relation network can effectively capture inter-instance correlations and thereby reduce the overall average bias. Notably, the MSE increase caused by removing c-RGT is substantially larger than that from removing s-RGT, indicating that leveraging cross-instance relations is more effective in mitigating extreme errors. The complementary roles of s-RGT and c-RGT together contribute to the optimal performance of the full RE-Rag model.
\begin{table}[!t]
\centering
\small
\caption{Ablation study of RE-Rag on the SMPD dataset.}
\label{tab:ablation}

\renewcommand{\arraystretch}{1.0}

\begin{tabularx}{\linewidth}{@{} l >{\raggedright\arraybackslash}X r r r @{}}
\toprule
\textbf{Module} & \textbf{Variant} & \textbf{MSE} & \textbf{MAE} & \textbf{SRC} \\
\midrule

Retriever & w/o Retrieval   & 1.5690 & 0.7923 & 0.8617 \\
          & w/o Attribute  & 0.9101 & 0.5041 & 0.9233 \\
          & w/o LR         & 0.7682 & 0.4855 & 0.9320 \\
          & w/o GR         & 0.7498 & 0.4762 & 0.9325 \\

\midrule

Modal     & w/o Cross       & 0.7514 & 0.4751 & 0.9339 \\
          & w/o Image      & 0.7631 & 0.4763 & 0.9331 \\
          & w/o Text       & 0.7843 & 0.4889 & 0.9306 \\

\midrule

\multirow[c]{3}{*}{\shortstack{Relation-Guided \\ Transformer}}
          & w/o Relation Map & 0.7801 & 0.4867 & 0.9314 \\
          & w/o s-RGT       & 0.7547 & 0.5125 & 0.9334 \\
          & w/o c-RGT       & 0.7756 & 0.5123 & 0.9340 \\

\midrule

\multicolumn{2}{l}{\textbf{RE-Rag (Full Model)}} 
& \textbf{0.7263} & \textbf{0.4577} & \textbf{0.9355} \\

\bottomrule
\end{tabularx}

\end{table}

\subsection{Parameter Sensitivity}

\subsubsection{\textbf{Number of retrieved instances $N_{retrieval}$}}
\begin{figure}[t]
    \centering
    \includegraphics[width=\columnwidth,keepaspectratio]{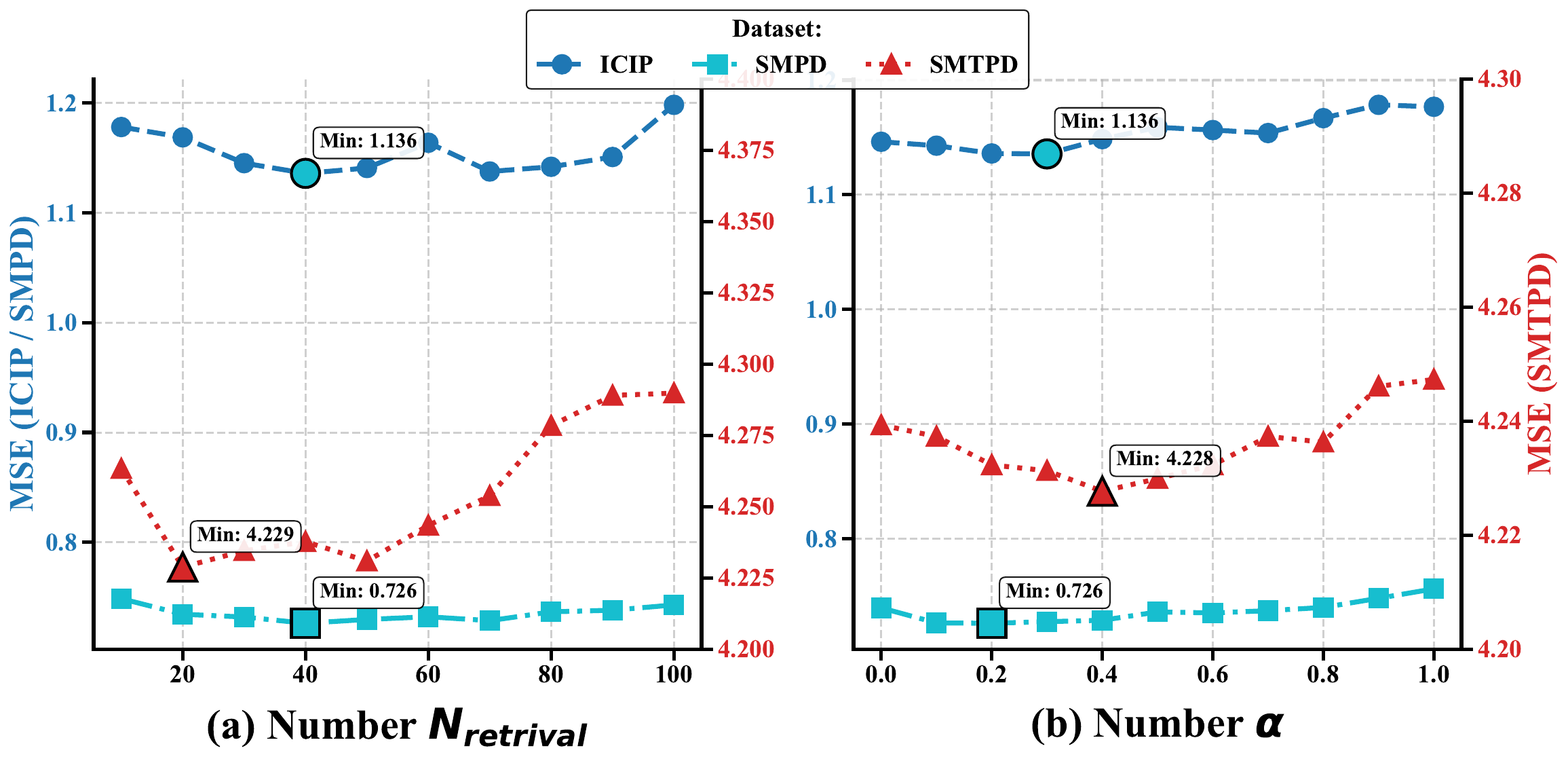}
    \caption{Parameter sensitivity analysis of RE-Rag on three datasets: (a) Number of retrieved instances $N_{retrieval}$, (b) Number of retrieval modality weight $\alpha$.}
    \label{fig:ak}
\end{figure}

Fig.~\ref{fig:ak} (a) illustrates the model’s performance across different retrieval set sizes $N_{retrieval}$. The results show that variations in the number of retrieved instances exert only a minor effect on the accuracy of the prediction. 
This robustness mainly benefits from the retriever's ability to prioritize the recall of key instances based on attribute cues. Even if the retrieval results contain some samples with relatively low relevance, the subsequent retrieval feature extraction module can effectively filter out noise through relative relationship modeling, thereby ensuring the stability and reliability of the prediction results. Based on this, we set $N_{retrieval}$ to 40 on all datasets to achieve a balance in overall performance.

\subsubsection{\textbf{Impact of Retrieval Modality Weighting.}}
Fig.~\ref{fig:ak}(b) shows the impact of retrieval modality weight $\alpha$ on model performance across different datasets, where $\alpha \in [0,1]$ denotes the weight assigned to the image modality. We observe that extreme values of $\alpha$ lead to degraded performance. When $\alpha$ is relatively small, the model generally performs well, indicating that textual information often carries more valuable features. Moreover, on the SMTPD dataset, the optimal value of $\alpha$ is slightly higher, which we attribute to platform differences. Considering the performance across datasets, we finally fix $\alpha$ to 0.3 to achieve a balanced setting for all datasets.

\begin{table*}[]
\centering
\renewcommand{\arraystretch}{1.2}
\begin{tabular}{|c|c|c|c|c|c|c|c|c|}
\hline
Target & Method & Top1  & Top2  & Top3 & Method & Top1 & Top2 & Top3 \\
\hline
\multirow[c]{7}{*}{\makecell[c]{\raisebox{-0.2\height}{\includegraphics[width=2.0cm]{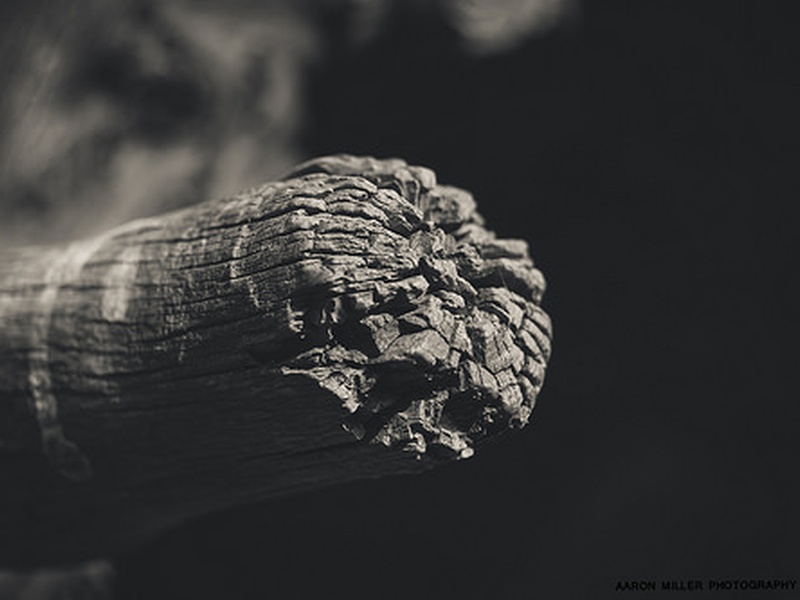}}}} &
  \multirow[c]{2}{*}{RE-Rag} &
  \makecell[c]{\raisebox{-0.2\height}{\includegraphics[width=1.6cm]{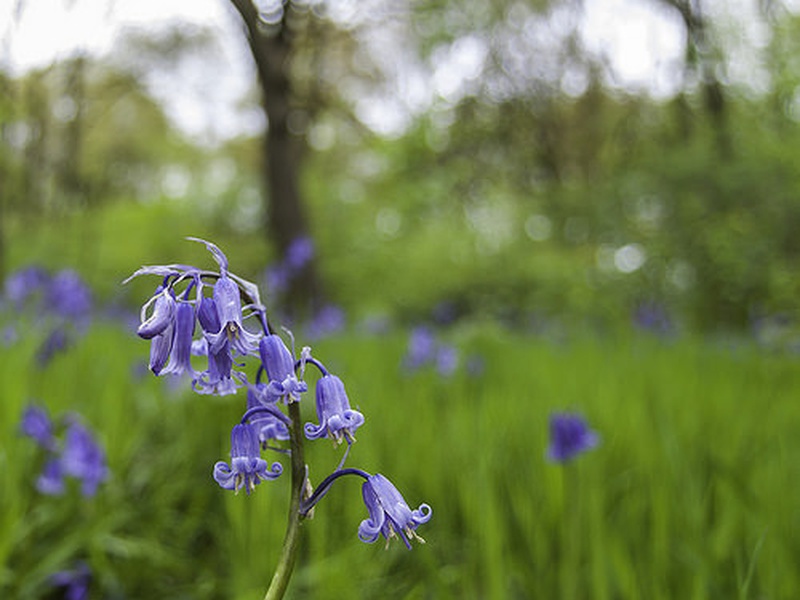}}}&
  \makecell[c]{\raisebox{-0.2\height}{\includegraphics[width=1.6cm]{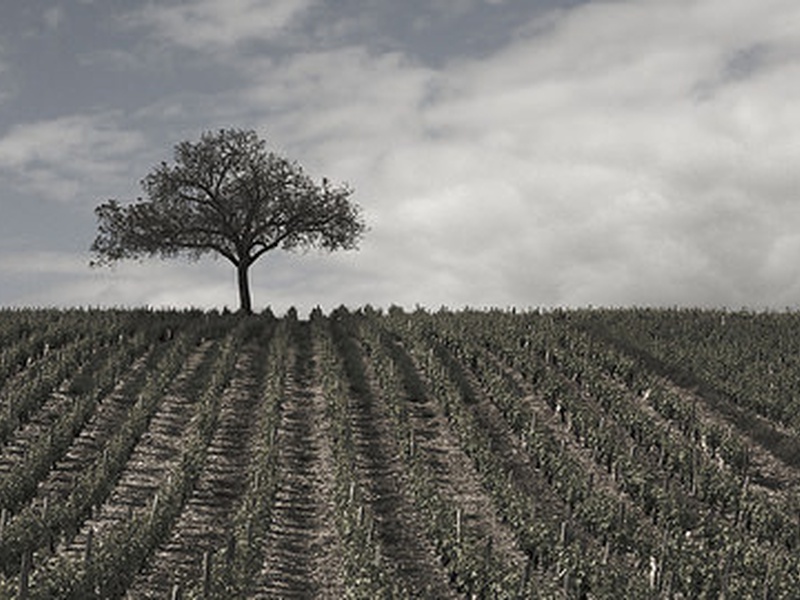}}} &
  \makecell[c]{\raisebox{-0.2\height}{\includegraphics[width=1.6cm]{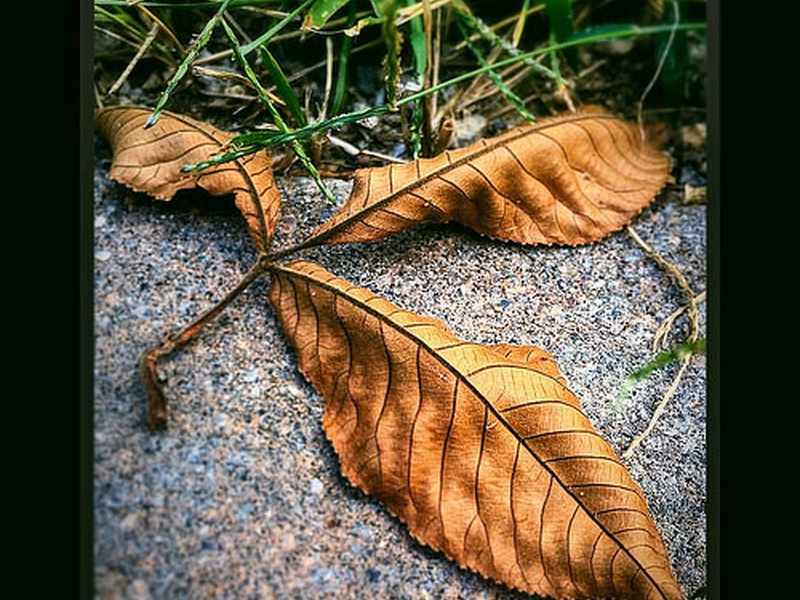}}} &
  \multirow[c]{2}{*}{MMRA} &
  \makecell[c]{\raisebox{-0.2\height}{\includegraphics[width=1.6cm]{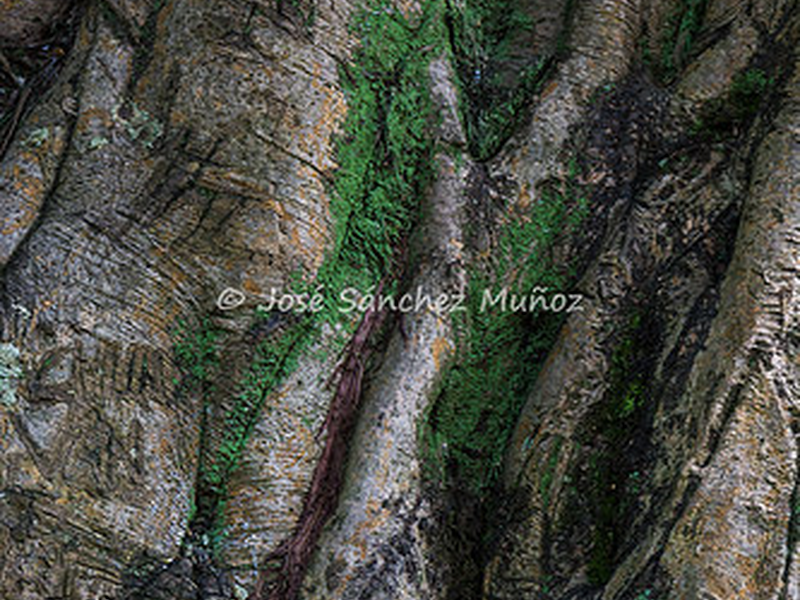}}} &
  \makecell[c]{\raisebox{-0.2\height}{\includegraphics[width=1.6cm]{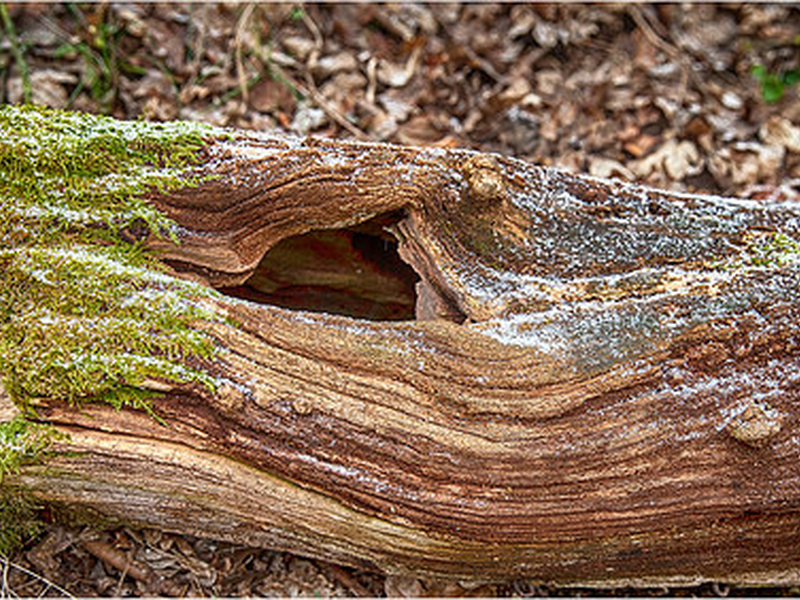}}} &
  \makecell[c]{\raisebox{-0.2\height}{\includegraphics[width=1.6cm]{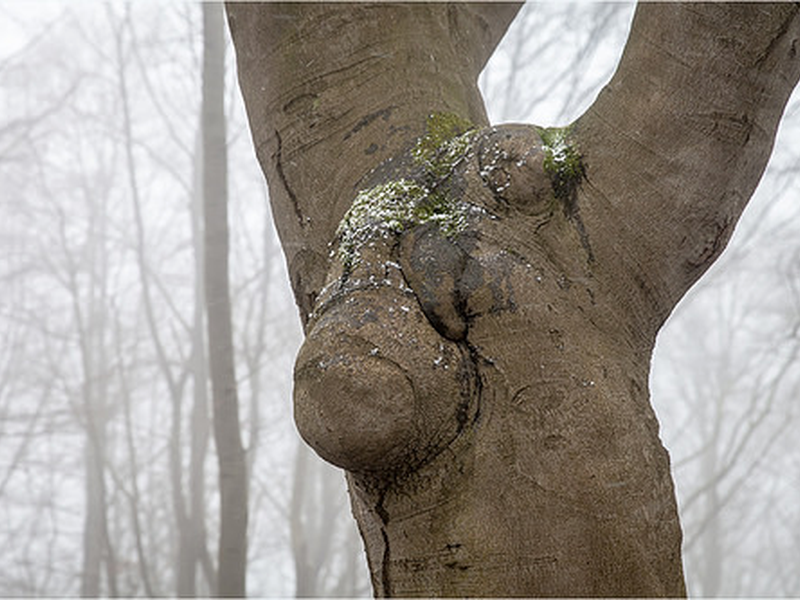}}} \\
\cline{3-5} \cline{7-9}
 &
   &
  \makecell[c]{Bluebell Wood} &
  \makecell[c]{Vine\\\#tree \#art} &
  \makecell[c]{Falling: Crisp} &
   &
  \makecell[c]{Old tree trunk} &
  \makecell[c]{Wood} &
  \makecell[c]{Trees \\\& Faces} \\
\cline{2-9}
       & 11.82     & 12.64 & 10.42 & 9.86 & 6.86     & 4.17 & 7.13 & 7.81 \\
\cline{2-9}
 &
  \multirow[c]{2}{*}{RAGTrans} &
  \makecell[c]{\raisebox{-0.2\height}{\includegraphics[width=1.6cm]{Img/q2t1.jpg}}} &
  \makecell[c]{\raisebox{-0.2\height}{\includegraphics[width=1.6cm]{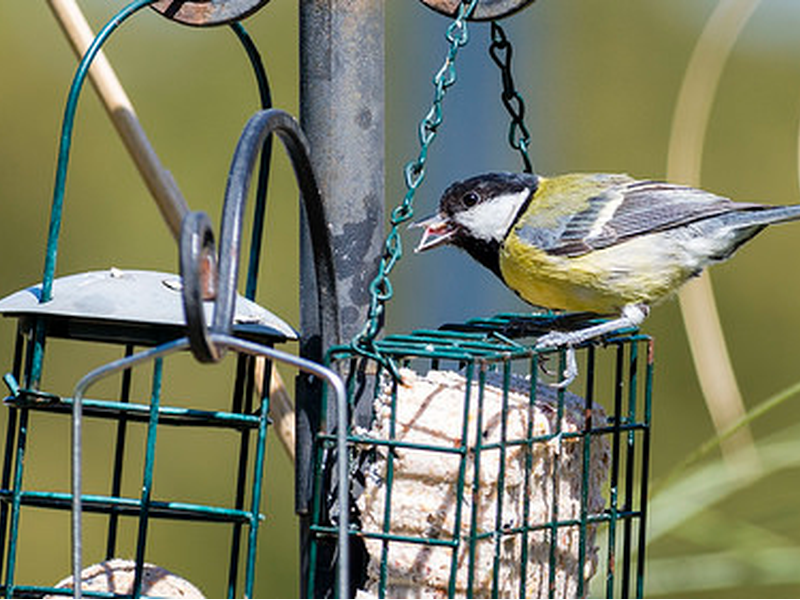}}} &
  \makecell[c]{\raisebox{-0.2\height}{\includegraphics[width=1.6cm]{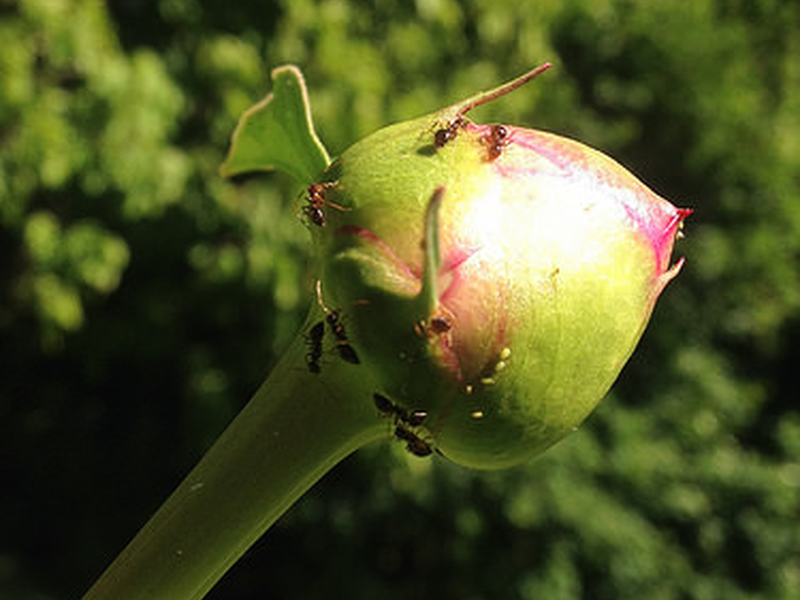}}} &
  \multirow[c]{2}{*}{SKAPP} &
  \makecell[c]{\raisebox{-0.2\height}{\includegraphics[width=1.6cm]{Img/q2t1.jpg}}} &
  \makecell[c]{\raisebox{-0.2\height}{\includegraphics[width=1.6cm]{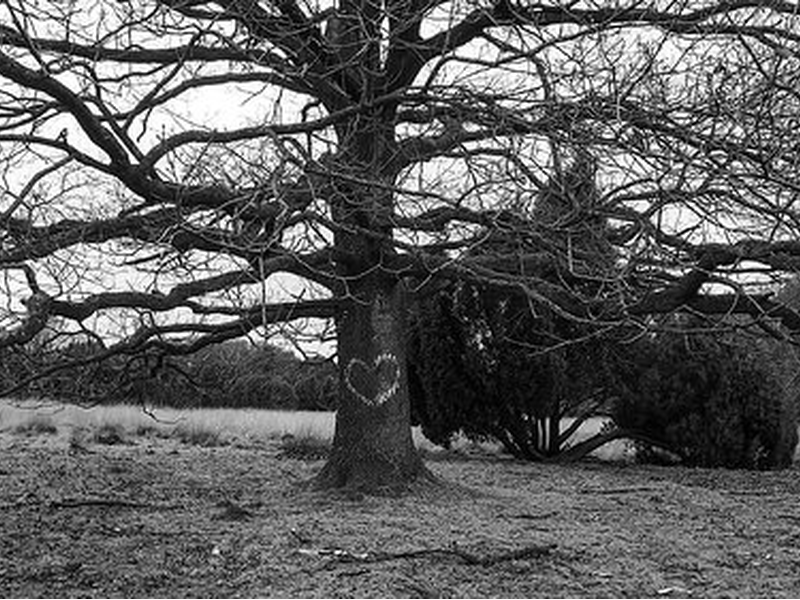}}} &
  \makecell[c]{\raisebox{-0.2\height}{\includegraphics[width=1.6cm]{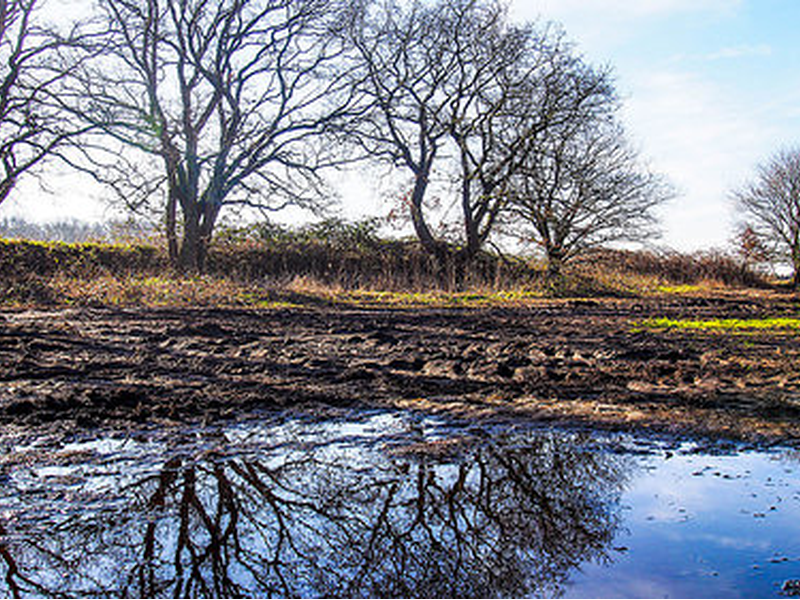}}} \\
\cline{1-1} \cline{3-5} \cline{7-9}
\makecell[c]{Branch end \\ \#camera \#trees} &
   &
  \makecell[c]{Bluebell Wood} &
  \makecell[c]{Adult \\Great Tit} &
  \makecell[c]{Insects \\lovenature} &
   &
  \makecell[c]{Bluebell Wood} &
  \makecell[c]{Reflection} &
  \makecell[c]{The Love \\tree} \\
\hline
12.51  & 11.03     & 12.64  & 7.32  & 9.06 & 10.33     & 12.64 & 8.79 & 8.98 \\
\hline
\end{tabular}
\caption{Top-3 Retrieved Instances and Popularity under Different Variants}
\label{tab:comparison}
\end{table*}

\subsection{Case Study}
\subsubsection{\textbf{Analysis of Retrieval Quality.}}

\begin{figure}[t]
    \centering
    \includegraphics[width=0.45\textwidth]{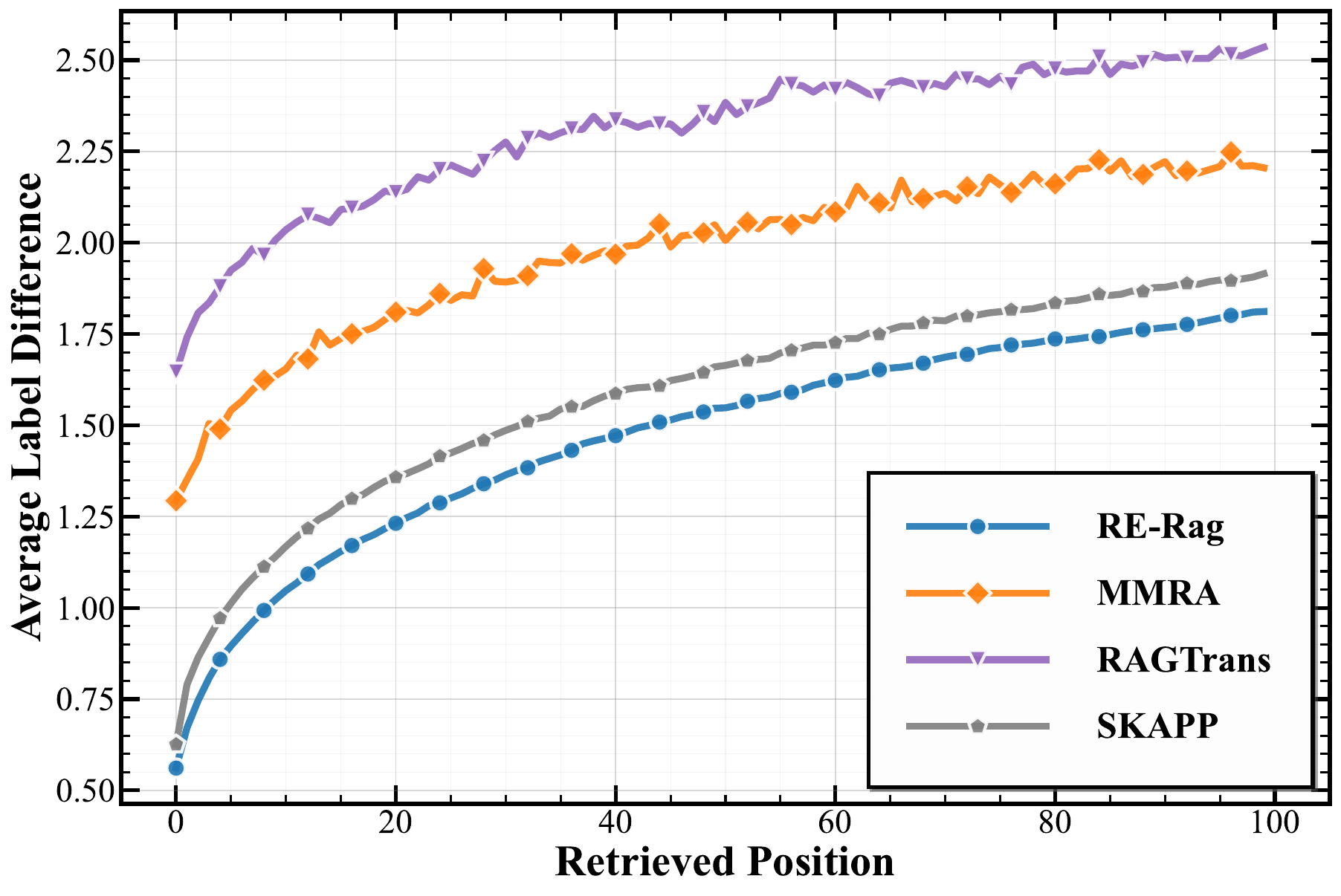}
    \caption{Average popularity gap between query samples and retrieved results across ranking positions on the SMPD.}
    \label{fig:re}
\end{figure}

\noindent To evaluate the effectiveness of retrieval quality, we compared three retrieval methods (MMRA, SKAPP and RAGtrans) and plotted the average popularity difference curves between the target instances and their top 100 retrieved results. This metric can, to some extent, reflect the retrieval model's performance in similarity discrimination and ranking precision. As shown in Fig.~\ref{fig:re}, RE-Rag effectively outperforms other baselines in terms of smaller average difference magnitude and more stable fluctuations, indicating its ability to discover samples with similar diffusion patterns while achieving finer-grained similarity discrimination and ranking. In contrast, RAGtrans exhibits the poorest performance, as it abstracts content semantics into aspect-level representations and incorporates attribute features via BM25-based weighting scheme. The lack of fine-grained interactions leads to limited discriminative power in retrieval results.

\subsubsection{\textbf{Analysis of retrieved instances.}}
We randomly select a UGC sample from the test set for retrieval visualization and compare it with several strong baseline retrieval-augmented models. As shown in Table~\ref{tab:comparison}, compared with other methods, our approach not only achieves broader coverage of retrieved instances but also exhibits closer consistency with the target sample in terms of popularity distribution, thereby validating the effectiveness of the Semantic-Attribute retriever.

\subsubsection{\textbf{Analysis of Retrieval Feature Extractor.}}
\begin{table}[!t]
\centering
\normalsize
\setlength{\tabcolsep}{8pt}
\renewcommand{\arraystretch}{0.9} 
\caption{Performance of Different Retrieval Feature Extractors under a Common Semantic-Attribute Retriever Input.}
\begin{tabular}{l r r r}
\toprule
\textbf{Variant} & \textbf{MSE} & \textbf{MAE} & \textbf{SRC} \\
\midrule
Prediction via MMRA \cite{MMRA} & 0.9549 & 0.5723 & 0.9103 \\
Prediction via SKAPP \cite{skapp} & 0.9721 & 0.5667 & 0.9197 \\
Prediction via RAGtrans \cite{RAGtrans} & 0.8551 & 0.5815 & 0.9240 \\
\textbf{RE-Rag(RGTs)} & \textbf{0.7263} & \textbf{0.4577} & \textbf{0.9355} \\
\bottomrule
\label{tab:re}
\end{tabular}
\end{table}
To further evaluate the effectiveness of the retrieval feature extractor, we conducted comparative experiments by replacing the retrieval feature extractor module of RE-Rag with the corresponding components from other retrieval-augmented paradigms (MMRA, SKAPP and RAGtrans), while keeping the retrieval input consistent. Specifically, MMRA adopts a bipolar attention mechanism, whereas RAGtrans leverages a hypergraph transformer to extract auxiliary features from retrieved instances for popularity prediction. As shown in Table~\ref{tab:re}, our proposed RGTs achieves the best performance under this setting, which confirms that modeling relative attribute relations effectively enhances the retrieval feature extractor.

\begin{table}[!t]
\centering
\normalsize 
\renewcommand{\arraystretch}{0.8} 
\caption{Complexity analysis on SMTPD.}
\begin{tabularx}{\columnwidth}{l *{3}{>{\raggedleft\arraybackslash}X}}
\toprule
\textbf{Model} & \textbf{Retr.} & \textbf{Inference} & \textbf{Param. size} \\
\midrule
MMRA & 1.6h & \textbf{2.2min} & 18.12M \\
ICPF & 2.9h & 2.0h & 213.97M \\
SKAPP & 6.5h & \underline{2.3min} & \textbf{13.62M} \\
RAGtrans & \underline{1.2h} & 3.4min & 44.24M \\
RE-Rag & \textbf{72s} & 2.8min & \underline{16.86M} \\
\bottomrule
\end{tabularx}
\label{tab:set}
\end{table}

\subsubsection{\textbf{Model robustness.}}
\begin{figure}[t]
    \centering
    \includegraphics[width=0.48\textwidth]{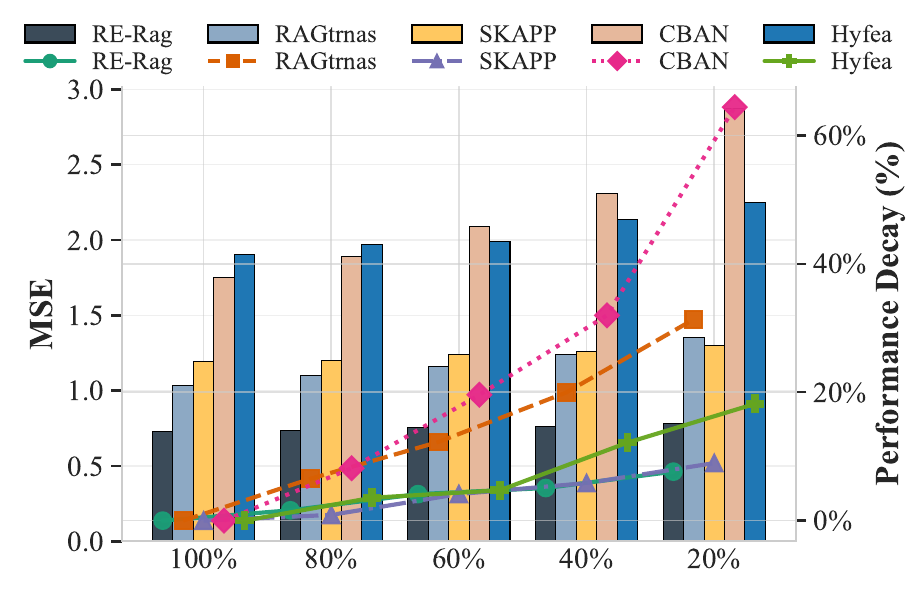}
    \caption{Impact of training set proportion on SMPD: bars indicate MSE, line indicates performance degradation gain.}
    \label{fig:robustness}
\end{figure}

We rigorously evaluated the robustness of RE-Rag against four typical baseline methods. Fig.~\ref{fig:robustness} illustrates the performance of multiple baselines under varying training set sizes. Notably, RE-Rag consistently achieves the highest performance across all settings, substantially outperforming the competing methods and demonstrating its exceptional stability and generalization capabilities.

\section{Conclusion}
In this study, we propose a Relationship-Enhanced Retrieval-Augmented Framework (\textit{RE-Rag}) for media popularity prediction. 
Its core idea is to model relative relationships among UGCs from a dual-factor perspective, considering both content quality and distribution mechanism, and incorporate them into the retrieval and prediction process to effectively leverage valuable historical instances.
This enhances the accuracy and generalization capability of popularity predictions. Extensive experiments on three real-world datasets validate the effectiveness and superior performance of this approach.

\section{Acknowledgments}
This work is supported by the Key R\&D Program of Zhejiang under Grant No.\ 2023C01044.

\bibliographystyle{ACM-Reference-Format}
\bibliography{arxiv}

\end{document}